\documentclass[twocolumn]{aastex631}
\usepackage{amsmath}
\submitjournal{AAS Journal}

\shorttitle{Astrometric Noise and Planetary Detection Efficiency due to Stellar Spots and Faculae}
\shortauthors{Chunhui Bao et al.}

\graphicspath{{./}{figures/}}

\begin{document}

\title{Closeby Habitable Exoplanet Survey (CHES). \uppercase\expandafter{\romannumeral1}. Astrometric Noise and Planetary Detection Efficiency due to Stellar Spots and Faculae}


\author{Chunhui Bao}
\affiliation{CAS Key Laboratory of Planetary Sciences, Purple Mountain Observatory, Chinese Academy of Sciences, Nanjing 210023, China;jijh@pmo.ac.cn}
\affiliation{School of Astronomy and Space Science, University of Science and Technology of China, Hefei 230026, China}

\author{Jianghui Ji}
\affiliation{CAS Key Laboratory of Planetary Sciences, Purple Mountain Observatory, Chinese Academy of Sciences, Nanjing 210023, China;jijh@pmo.ac.cn}
\affiliation{School of Astronomy and Space Science, University of Science and Technology of China, Hefei 230026, China}
\affiliation{CAS Center for Excellence in Comparative Planetology, Hefei 230026, China}

\author{Dongjie Tan}
\affiliation{CAS Key Laboratory of Planetary Sciences, Purple Mountain Observatory, Chinese Academy of Sciences, Nanjing 210023, China;jijh@pmo.ac.cn}
\affiliation{School of Astronomy and Space Science, University of Science and Technology of China, Hefei 230026, China}

\author{Guo Chen}
\affiliation{CAS Key Laboratory of Planetary Sciences, Purple Mountain Observatory, Chinese Academy of Sciences, Nanjing 210023, China;jijh@pmo.ac.cn}
\affiliation{School of Astronomy and Space Science, University of Science and Technology of China, Hefei 230026, China}

\author{Xiumin Huang}
\affiliation{CAS Key Laboratory of Planetary Sciences, Purple Mountain Observatory, Chinese Academy of Sciences, Nanjing 210023, China;jijh@pmo.ac.cn}
\affiliation{School of Astronomy and Space Science, University of Science and Technology of China, Hefei 230026, China}

\author{Su Wang}
\affiliation{CAS Key Laboratory of Planetary Sciences, Purple Mountain Observatory, Chinese Academy of Sciences, Nanjing 210023, China;jijh@pmo.ac.cn}
\affiliation{CAS Center for Excellence in Comparative Planetology, Hefei 230026, China}

\author{Yao Dong}
\affiliation{CAS Key Laboratory of Planetary Sciences, Purple Mountain Observatory, Chinese Academy of Sciences, Nanjing 210023, China;jijh@pmo.ac.cn}
\affiliation{CAS Center for Excellence in Comparative Planetology, Hefei 230026, China}

\begin{abstract}

    The Closeby Habitable Exoplanet Survey (CHES) is dedicated to the astrometric exploration for habitable-zone Earth-like planets orbiting solar-type stars in close proximity, achieving unprecedented micro-arcsecond precision. Given the elevated precision, thorough consideration of photocenter jitters induced by stellar activity becomes imperative. This study endeavors to model the stellar activity of solar-type stars, compute astrometric noise, and delineate the detection limits of habitable planets within the astrometric domain. Simulations were conducted for identified primary targets of CHES, involving the generation of simulated observed data for astrometry and photometry, accounting for the impact of stellar activity. Estimation of activity levels in our samples was achieved through chromospheric activity indices, revealing that over 90\% of stars exhibited photocenter jitters below 1 $\mu\mathrm{as}$. Notably, certain proximate stars, such as $\alpha$ Cen A and B, displayed more discernible noise arising from stellar activity. Subsequent tests were performed to evaluate detection performance, unveiling that stellar activity tends to have a less pronounced impact on planetary detectability for the majority of stars. Approximately 95\% of targets demonstrated a detection efficiency exceeding 80\%. However, for several cold stars, e.g., HD 32450 and HD 21531, with the habitable zones close to the stars, a reduction in detection efficiency was observed. These findings offer invaluable insights into the intricate interplay between stellar activity and astrometric precision, significantly advancing our understanding in the search for habitable planets.

\end{abstract}


\section{Introduction} \label{sec:intro}
Over several decades, the discovery of more than 5500 exoplanets has been achieved through the application of diverse methods \footnote[1]{\url{https://exoplanetarchive.ipac.caltech.edu}}. The predominant techniques employed for these discoveries are the transit and radial velocity (RV) methods. However, these two approaches are susceptible to the impact of stellar activity, capable of inducing RV variations akin to those attributed to orbiting planets. With the development of the new generation of extreme-precision spectrographs, the radial-velocity (RV) measurement at sub-meter-per-second precision becomes feasible, including the Echelle SPectrograph for Rocky Exoplanets and Stable Spectroscopic Observations (ESPRESSO) \citep{2010ESPRESSO}, the EXtreme PREcision Spectrograph (EXPRES) \citep{2016EXPRES} and NEID \citep{2022Lin}. The previous investigations showed that the amplitude of RV variations resulting from stellar activity ranges in the order of a few $\mathrm{m/s}$ \citep{2002Hatzes, 2010Lagrange, 2006Butler, 2017Butler, 2023Laliotis}. The dispersion in radial velocity due to granulation is estimated to be larger than $0.3~\mathrm{m/s}$, significantly surpassing the signals emanating from terrestrial planets (approximately $\sim 0.09~\mathrm{m/s}$). \citet{2022Zhao} reported that EXPRES is unable to detect Earth-like planets in the presence of stellar variability. Periodic variability in stellar activity can mimic a planetary transit signal, leading to potential false positives \citep{2009Lanza}. Photometry observations are also influenced by stellar oscillations and granulation, which are comparable to the planetary signal at the precision of CHEOPS (CHaracterising ExOPlanet Satellite) and ESPRESSO \citep{2023Sulis}. The simulations of PLATO (PLAnetary Transits and Oscillations of stars) concluded that the noise due to granulation is about 100 ppm \citep{2020Morris}, comparable to the depth of Earth-like planets.  Despite efforts to bolster the performance of the RV method through denoising techniques such as Gaussian processes and extended observational durations (spanning ten years), uncertainties in mass estimations persistently exceed 30\% \citep{2023Meunier}.

The astrometric method offers distinct advantages in the detection of terrestrial planets. Several astrometric missions, among them the Gaia satellite, have made exoplanet detection a primary scientific objective. Notably, the discovery of over 60 Jupiter-mass planets through astrometry was reported in Gaia DR3 \citep{2023Gaia}. Numerous upcoming missions are specifically dedicated to the search for Earth-like planets, including Theia mission \citep{2017Theia}, the Closeby Habitable Exoplanet Survey (CHES) mission \citep{2022Ji}, Small-JASMINE mission \citep{2023Kawata}, and the Nancy Grace Roman Telescope \citep{2023Gandhi}, Habitable Worlds Observatory (HWO) survey \citep{2024Mamajek}. For instance, Theia endeavors to identify approximately six habitable terrestrial planets, a breakthrough with profound implications for understanding the conditions conducive to planetary formation and the emergence of life \citep{2017Theia}.

The meticulous consideration of stellar activity noise proves indispensable when operating at ultra-high precision. Predominantly, the primary astrometric shifts arise from the intricate interplay of stellar spots and faculae. Stellar spots, magnetic structures that manifest as darkened regions on the stellar surface, comprise a darker core called the umbra, surrounded by a slightly bright halo known as the penumbra \citep{2003Solanki}. Faculae, existing in proximity to spots, exhibit a slightly higher temperature than the photosphere. Consequently, while facular contrast in temperature is less pronounced compared to spots, their significantly larger areas contribute crucially to astrometric shifts.

Considerable investigation has been dedicated to the intricate task of modeling astrometric jitter induced by spots, particularly in anticipation of the Space Interferometry Mission (SIM) \citep{2007Eriksson, 2009Makarov, 2010Makarov}. Their findings suggested the feasibility of detecting Earth-like planets around quiescent stars analogous to the Sun. However, challenges arise when dealing with more active stars. \citet{2011Lagrange} extended their considerations to encompass both plages and spots, utilizing solar spots groups and bright structures from USAF/NOAA and MDI/SOHO and revealed that the amplitude of activity-induced astrometric signals remains below 0.2 $\mu\mathrm{as}$.

Based on their model, \citet{2019Meunier} incorporated granulation and chromospheric emission, extending the spectral type to F6-K4 stars. They determined that the root mean square (rms) of the astrometric activity time series for other main-sequence stars can be two to five times the solar value. Nevertheless, the detection rates for habitable terrestrial planets persist above 50\%, marking a substantial improvement compared to the RV method \citep{2020Meunier}. In their analysis of simulated time series, \citet{2022Meunier} concentrated on 55 nearby stars in the Theia sample. Their findings echoed similar conclusions, suggesting that stellar activity minimally impacts the detection of planets for solar-type stars, except for the closest stars $\alpha$ Cen A and B. They also emphasized the need to enhance the observation strategy for certain subgiants due to the more distant habitable zone. Additionally, studies based on Gaia accuracy, such as \citet{2018Morris}, developed a model attributing stellar jitter to spots and concluded that Gaia's precision is adequate to measure photocenter jitter for some M dwarfs within 10 $\mathrm{pc}$.

In a recent study, \citet{2021Shapiro} computed the displacements observable in both the Gaia and Small-JASMINE passbands \citep{2023Kawata}. Notably, they identified a jitter amplitude in the G-band of approximately 0.5 $\mu\mathrm{as}$ for Sun located at 10 $\mathrm{pc}$, anticipated to be twice as large as that in the Small-JASMINE near-infrared passbands. Expanding their investigation to stars observed at diverse inclinations, metallicity levels, and active-region nesting degrees \citep{2021Sowmya}, they discovered that astrometric shifts peak at an intermediate inclination in the Gaia-G band. Building upon the understanding that active magnetic structures tend to emerge in higher latitude regions \citep{2009Strassmeier}, \citet{2022Sowmya} advanced their research by exploring stars with solar fundamental parameters but faster rotation than the Sun. Their findings indicated that magnetic jitter is predominantly spot-dominated for rapid rotators. In an exploration of the multi-wavelength strategy in astrometry, \citet{2022Kaplan} employed solar simulations across five distinct passbands based on \citet{2021Shapiro}. Their results highlighted that the minimum detectable mass, through a combination of observations like Gaia G and R bands, is 0.005 Earth mass. However, it is crucial to note that their assumptions involved perfect instruments, which can introduce much larger noise than stellar activity in practical scenarios.

The primary goal of this work is to evaluate the impact of stellar activity on the astrometric signal of the target stars of CHES. We developed a model that incorporates spots and faculae to simulate stellar activity in astrometry. The stellar activity levels were derived from photometric data obtained from TESS and the chromospheric index $\mathrm{log} R^\prime_{\mathrm{HK}}$. Subsequently, we generated simulated astrometric time series for each target. Noteworthy, we observed that nearly 75\% of stars exhibit minimal photocenter jitter, with standard deviations lower than 0.24 $\mu \mathrm{as}$. Moreover, we introduced simulated habitable planets and conducted retrieval experiments to ascertain their mass and period. The retrieval performance proved to be satisfactory, with approximately 80\% of simulated planets around nearly 95\% of the stars being successfully recovered in our assessments. These findings underscore the feasibility of the astrometry method for detecting terrestrial planets, even when considering the noise introduced by stellar activity.

The paper is structured as follows: Section \ref{sect:Obs} provides a brief description of the properties of targets and planets. In Section \ref{sect:method}, we elaborate on the stellar activity model, delineate the properties of spots and faculae, and elucidate the procedure for generating simulated stellar jitter at each time step. The methodology employed to estimate the detection efficiency of stars is outlined in Section \ref{sect:detectability}. Subsequently, Section \ref{sect:result} presents the outcomes of simulations and the detection efficiency. In the final section, we summarize the major results.

\section{Targets and approach}
\label{sect:Obs}

\subsection{CHES satellite}
The Closeby Habitable Exoplanet Survey (CHES) mission is set forth to unveil habitable-zone Earth-like planets surrounding solar-type stars in close proximity, approximately 10 parsecs distant from our solar system with micro-arcsecond relative astrometry techniques. The primary scientific objectives of CHES encompass the detection of Earth Twins or terrestrial planets within habitable zones orbiting 100 FGK nearby stars \citep{2022Ji}. Moreover, the mission aims to execute an exhaustive survey and intricately characterize the neighboring planetary systems, and reveal the real planetary masses and 3-dimensional orbits that provide crucial clues to their planetary formation and dynamical evolution \citep{2005Youdin, 2007Johansen, 2016Jin, 2019Huang, 2019Liu, 2022Pan, 2024Pan}. CHES is expected to enrich the sample of Earth-like planets, which helps to comprehensively understand the formation and evolution of planetary systems \citep{2004Ida, 2012Mordasini, 2014Jin, 2020Liu, 2022Huang, 2023Huang}. The CHES satellite, outfitted with a 1.2-meter aperture high-quality telescope featuring low distortion and high stability, will be stationed at the Sun-Earth L2 point, continuously observing all target stars over a 5-year duration \citep{2022Ji}. Within a planetary system, the host star undergoes a subtle wobble around the common center of mass, typically demonstrating an amplitude of 0.3 $\mu \mathrm{as}$, which is measured for a solar-mass star situated at a distance of 10 parsecs, influenced by an Earth-mass planet located at 1 $\mathrm{AU}$. In the astrometry, our measurements are focused on the position of the stellar photocenter, operating under the assumption that it coincides with the center of mass. However, beyond the gravitational influence of planets, the stellar photocenter can experience displacement due to the magnetic activity of the host star. Bright faculae regions contribute to increased brightness and attract the photocenter, while dark spots exert the opposite effect. The resulting photocenter jitters are measured in milli stellar radii, a scale comparable to the astrometric signature of an Earth-like planet \citep{2008Catanzarite}.

\subsection{Target Selection}

\begin{figure}
   \centering
   \includegraphics[scale=0.55]{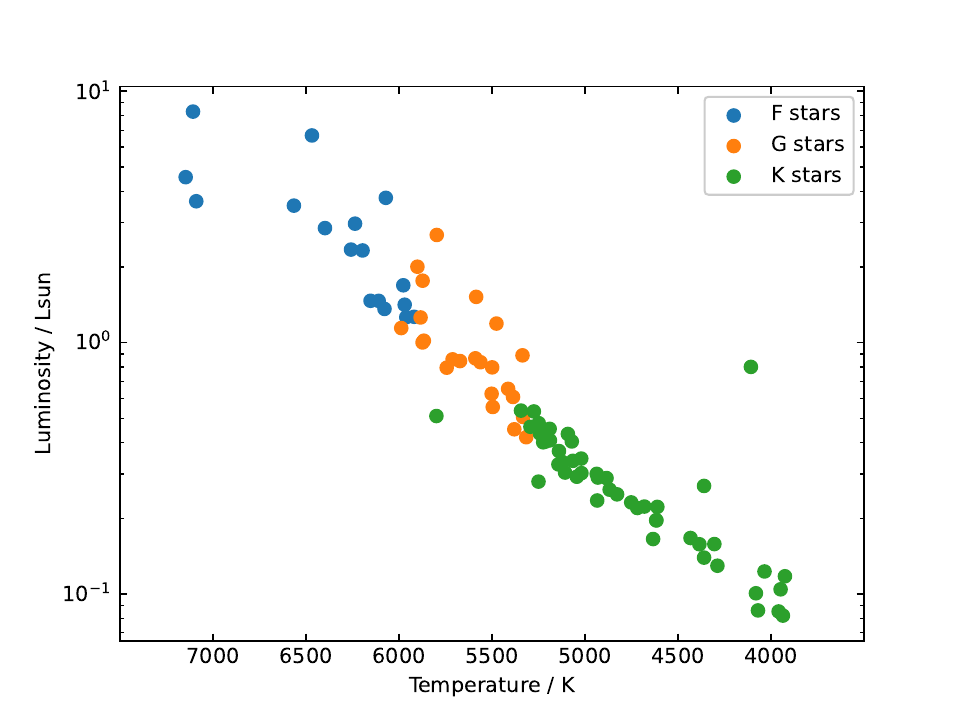}
   \caption{The Hertzsprung-Russell diagram depicts CHES targets, comprising 19 F stars, 24 G stars, and 51 K stars.}
   \label{HRD}
\end{figure}

In this study, we have identified 94 main sequence stars from the CHES targets catalogue. The distribution of effective temperature and luminosity is illustrated in Figure \ref{HRD}, with colors denoting the stellar types (e.g., F, G, K). Remarkably, the majority of these stars are positioned within approximately 10 $\mathrm{pc}$ from the solar system. Additionally, the astrophysical parameters of the target stars are derived primarily from the TIC8.2 input catalogue \citep{2021Paegert} and are succinctly presented in the appendix.

\subsection{Planetary signals}
The primary scientific goal of CHES is to detect the habitable planets orbiting nearby solar-type stars by observing the stellar wobbles induced by planetary gravity. The amplitude of the astrometric signal produced by a planet can be described as \citep{2014Perryman}:
\begin{eqnarray}
   \alpha &=&  \frac{M_p}{M_*} \, \frac{a}{D}  \nonumber  \\
       &=& 0.3 \left( \frac{M_p}{M_\oplus} \right) \left( \frac{a}{1 \mathrm{AU}}\right) \left( \frac{M_*}{M_\odot}\right)^{-1} \left(\frac{D}{10 \mathrm{pc}}\right)^{-1}  \mu \mathrm{as},
\label{alpha}
\end{eqnarray}
where $M_p$ denotes the planetary mass, $M_*$ is the stellar mass, $a$ represents the semi-major axis of the planet, and $D$ is the stellar distance to the solar system. Consequently, the astrometric signal induced by an Earth-mass planet at 1 au, causing a solar-mass star to wobble 10 $\mathrm{pc}$ away, is measured at 0.3 $\mu \mathrm{as}$ as aforementioned.

\section{Stellar activity model}
\label{sect:method}

We formulated a model to simulate the emergence of spots and faculae on solar-type stars. Subsequently, we generated synthetic observational data, encompassing both photometry and astrometry, spanning a five-year duration to match the expected time baseline of CHES.

\subsection{Analytic centroid approximation}

The difference in intensity of stellar magnetic features, including spots and faculae, compared to the photosphere, leads to variations in the stellar photocenter as the star rotates and features evolve. To evaluate the influence of jitters on the detection of terrestrial planets, we constructed a model employing {\tt\string butterpy} \citep{2022Claytor} to simulate the presence of spots and faculae. Subsequently, we computed the astrometric time-series.

We considered a star with a radius $R_*$ and defined Cartesian coordinates on the stellar surface to calculate variations in the photocenter across the stellar surface \citep{2018Morris}. The origin is placed at the center of the star, with the x-axis aligned along the stellar equatorial plane. The y-axis indicates the rotation axis, and the sign of the z-axis determines whether the feature is obstructed by the star. Thus, the position of the $i$th feature can be described as $\boldsymbol{r_i} = (x_i, y_i)$, where $r_i = \vert \mathbf{r_i} \vert$.

The total flux from the star is characterized by \citep{2018Morris}
\begin{eqnarray}
F_* = \int_0^{R_*} 2 \pi I(r) \mathrm{d} r ,
\end{eqnarray}
where $r$ is the distance to origin in unit of stellar radius, and $I(r)$ is limb-darkening law. We use a four-parameter the non-linear limb-darkening law introduced by \citet{2000Claret},
\begin{equation}
\frac{I(\mu)}{I(1)} = \left\{
\begin{aligned}
1 - \sum_1^4 a_k (1-\mu^{k/2}) \quad  \mu > \mu_{cri} \\
0 \quad \mu < \mu_{cri} ,
\end{aligned}
\right.
\end{equation}
where $\mu$ represents the position on the surface (cosine between the local surface and the line of sight), $\mu = \sqrt{1 - r^2}$, {$\mu_{cri}$ is the drop-off point, $a_k$ are limb-darkening coefficients}, which are linked to stellar effective temperature and log$g$ \citep{2018Claret}.

We generated a bulk of spots and faculae in the stellar surface, and the flux contribution from each feature is determined by its position, area, and intensity. The intensity of the $i$th spot can be described as \citep{2019Meunier}:
\begin{eqnarray}
    F_{\mathrm{sp},i} = - C_{\mathrm{sp}, i} A_i * \mu_i ,
\end{eqnarray}
where $C_{\mathrm{sp}, i}$ is the contrast of spot, $A_i$ is the area of spot.

Here we assume that all features are circular, with a radius $R_i$, a position $r_i$ and a true area $A_{0}$. Features near limb will be foreshortened, approximated by an ellipse with a semi-major axis $R_i$ and a semi-minor axis $R_i \sqrt{1-r_i^2}$. Thus the projected area can be expressed as:
\begin{equation}
    A_i = \sqrt{1-r_i ^2} A_0.
\end{equation}
The projected area is independent of $R_i$, so we only concern on true area and position of spots or faculae.

In this study, the temperature contrast $C_{\mathrm{sp}}$ is utilized. We assume that both the photosphere and the spot exhibit blackbody characteristics. Therefore, $C_{\mathrm{sp}}$ is described through the Planck function $f$ \citep{2019Meunier}:
\begin{eqnarray}
    C_{\mathrm{sp}} = 1 - \frac{f(T_{\mathrm{sp}}) I(\mu_i)}{f(T_{eff}) I(\mu_i)} .
\end{eqnarray}
And the intensity of faculae is similar as spots:
\begin{eqnarray}
    F_{\textrm{fa}, i} = A_i C_{\textrm{fa}, i} * \mu_i .
\end{eqnarray}

The astrometric contribution in the $x$ direction is the same for faculae and spots:
\begin{eqnarray}
   \begin{aligned}
    \Delta x &=  \frac{R_*}{D}\sum \frac{F_i}{F_*} x_i \\
            &= 0.465 \times 10^{2} \ \left(\frac{R_*}{R_\odot} \right) \left(\frac{D}{10 \mathrm{pc}}\right)^{-1} \sum \frac{F_i}{F_*} x_i \quad\mu\mathrm{as} ,
\label{astro-st}
   \end{aligned}
\end{eqnarray}
the formula in $y$ direction is similar.

\subsection{Fundamental stellar parameters}\label{s32}
In our model, the astrometric precision is directly influenced by the stellar radius in Equation \ref{astro-st}, while the planetary signal is connected to the stellar mass in Equation \ref{alpha}. In addition to mass and radius, the activity level of star determines numbers and distribution of magnetic features. In this paper, we adopt the dimensionless parameter activity level defined by \citet{2015Aigrain}, with the assumption that the Sun's level is 1. This parameter will be discussed further in section \ref{sect:level}.

Subsequently, our objective is to determine temporal variations in stellar activity, encompassing both the rotation period and cycle period. Rotation is a common observation in many stars, with typical periods ranging from a few days to a few hundred days. Due to the planned uneven sampling strategy during CHES observations \citep{2022Ji}, the shortest cadence is about two or three days, and short-term rotation also impacts astrometry. The rotation rate of each target is yet to be determined for simulating. {Previous studies have deduced stellar rotation periods from lightcurves or spectral observations \citep{2014McQuillan, 2020Pan, 2023Jin}, revealing a correlation between rotation period and characteristics in large star samples from TESS or Kepler \citep{2020Witzke, 2020Reinhold}.}

\begin{figure}
   \centering
   \includegraphics[scale=0.50]{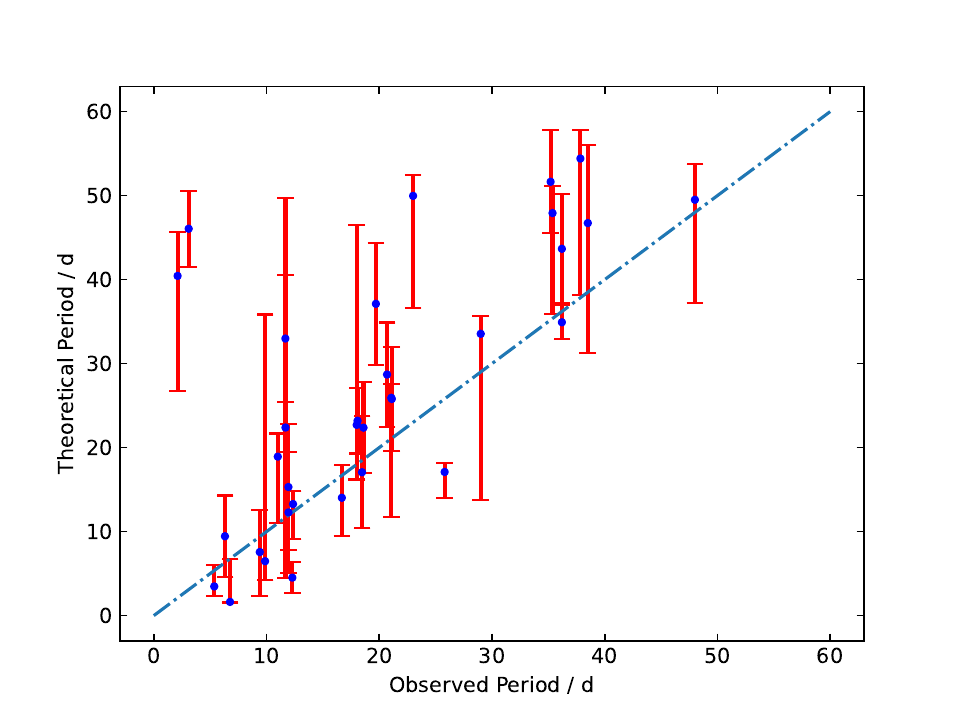}
   \caption{Theoretical and observed rotation periods of 33 stars in our sample, the dashed line denotes the observed periods equal to theoretical periods. Errorbars indicates the period ranges from uncertainties of $\mathrm{log} R^\prime_{\mathrm{HK}}$.}
   \label{periods}
\end{figure}

We have chosen to utilize the empirical law that establishes a relationship between the Rossby number $R_o$ and the estimated turnover timescale $\tau_c$ with the rotation period. The rotation period can be estimated according to \citet{1984Noyes}:
\begin{eqnarray}\label{equ_rot}
    P_{rot} = R_o \times \tau_c ,
\end{eqnarray}
where the Rossby number is estimated by $\mathrm{log} R^\prime_{\mathrm{HK}}$ \citep{2008Mamajek}, and the turnover timescale is linked to B-V \citep{2016Surez}. Figure \ref{periods} presents the theoretical and observed rotation period of 33 stars in our samples. The observed periods closely align with theoretical values, falling within the margin of error for roughly 80\% of the stars. Despite the more substantial disparity between the two stars with the shortest periods, our attention is directed towards stars characterized by moderate periods. This arises from the fact that only approximately 10\% of FGK stars manifest rotation periods of less than five days in the Kepler and K2 samples \citep{2014McQuillan, 2020Reinhold}. Consequently, we derived rotation periods for stars lacking observed periods with Equation \ref{equ_rot}.

Stellar activity cycles are widely hypothesized to arise from dynamo processes. The solar cycle spans approximately 11 years, and there is evidence suggesting a correlation between activity cycle periods and rotation periods \citep{1984Noyes, 1998Brandenburg}. In our study, we employed a relationship tailored for FGK stars to estimate the cycle period \citep{2023Irving}. The cycle period and rotation period details are presented in the appendix. All target cycle periods are shorter than five years (the time baseline of CHES), enabling us to assume that the maximum stellar activity jitter in a cycle is estimated in our simulations.

\subsection{Determination of activity level}
\label{sect:level}
Given that the activity level serves as an adjustable scale factor governing the average rate of spot and facula emergence \citep{2015Aigrain}, it becomes imperative to ascertain the activity level for each target. We determine the activity level using the chromospheric index $\mathrm{log} R^{\prime}_{HK}$ and photometric variability as detailed below.

Initially we derived the lightcurves of 78 stars observed by TESS by the package {\tt\string lightkurve} \citep{2018Lightkurve}, encompassing the simple aperture photometry (SAP Flux) and the presearch data conditioning SAP flux (PDCSAP Flux). For both of type data, we removed outlier and bad data based on quality flag of TESS. Each light curve was scrutinized visually to avoid systematic errors. And the SAP Flux is preferred as the long-term trends are removed in PDCSAP Flux. Subsequently, the peak-to-peak amplitude values were derived to characterize variability of each lightcurve \citep{2020Hojjatpanah}. Given that the timescale of a single sector is much shorter than the stellar cycle period, for stars observed in two or more sectors, we individually processed each light curve and calculated the peak-to-peak values. {In most cases, the differences in peak-to-peak values between different sectors are quite small, and the mean peak-to-peak values are derived as input parameters in our model. For individual stars, we found that the peak-to-peak value from one sector is significantly higher than the others, and no systematic errors were found. In this case, the lightcurve of this sector is considered to have been observed observed during the more active phase of the star, and the mean value of PDCSAP Flux and SAP Flux of was used.} Additionally, we calculated the variations of Sun from 25 Feb. 2003 to 25 Feb. 2020 as a comparison.

\begin{figure*}
   \gridline{\fig{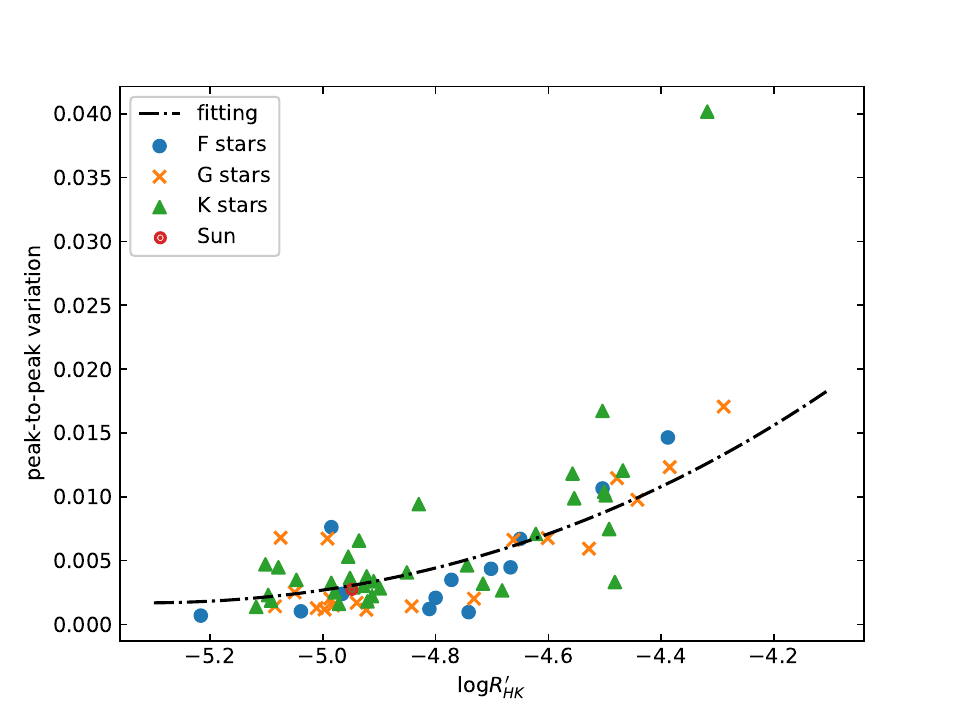}{0.45\textwidth}{(a)}
            \fig{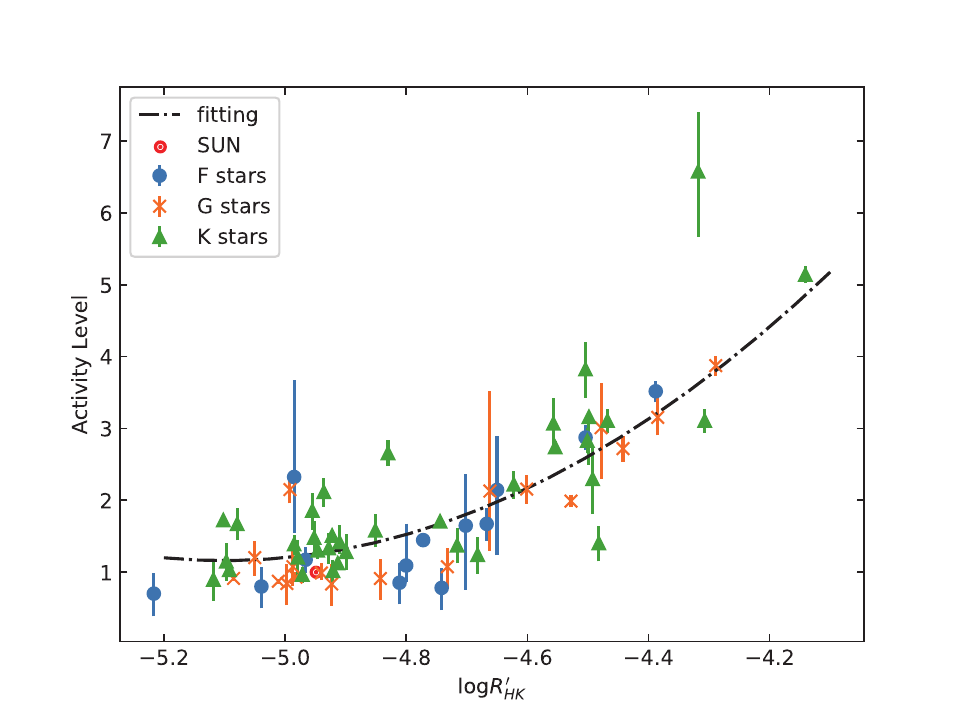}{0.45\textwidth}{(b)}}
   \caption{\emph{Left panel}: the relation between $\mathrm{log} R^{\prime}_{HK}$ and peak-to-peak variations of photometry, the dashed line shows a parabolic fitting. \emph{Right panel}: same for the relation between $\mathrm{log} R^{\prime}_{HK}$ and activity level, {while errorbars shows the uncertainties of activity level}. The blue circle, orange cross, and green triangle correspond to F,G,K stars respectively.
   \label{logR2level-p2p}}
\end{figure*}

\begin{figure}
   \centering
   \includegraphics[scale=0.50]{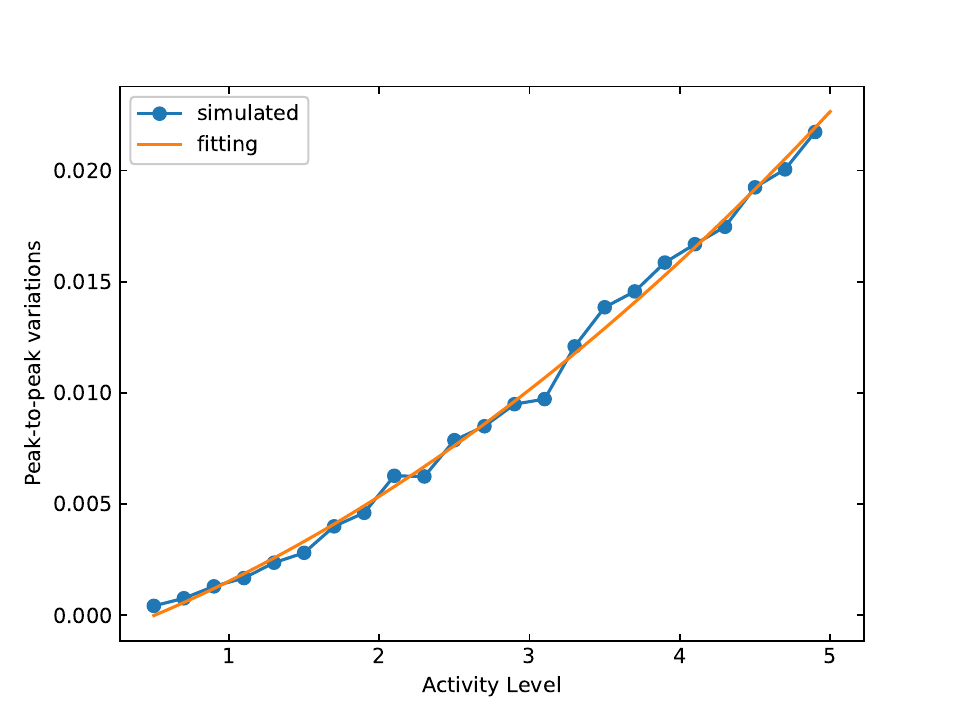}
   \caption{The activity level with photometric peak-to-peak variations. The step of activity level is 0.2, the y-coordinate of each point indicates the mean peak-to-peak variations of 100 simulated lightcurves.}
   \label{level-p2p}
\end{figure}

We identified a quadratic relation between peak-to-peak values and $\mathrm{log} R^{\prime}_{HK}$, {which represents the contribution of the CaII H and K lines to the bolometric luminosity of the star \citep{1984Noyes}. This index can be derived through spectral observations, such as those obtained by instruments like LAMOST \citep{2018Zong}.} Figure \ref{logR2level-p2p} displays $\mathrm{log} R^{\prime}_{HK}$ and peak-to-peak variability of 78 samples and Sun (left panel). The variation of the Sun was calculated based on Total Solar Irradiance(TSI) data \footnote[1]{TSI data is from observation of The Solar Radiation and Climate Experiment (SORCE)}. Photometric variations were utilized to relate $\mathrm{log} R^{\prime}_{HK}$ activity level in our model. We established a grid of levels from 0.5 to 5, and generated 100 simulated lightcurves at each level. Figure \ref{level-p2p} reveals a parabolic relationship {between activity level and peak-to-peak variation}, consistent with the results of \citet{2015Aigrain}. Consequently, activity levels of targets with observed lightcurves could be determined.

For stars not observed by TESS or Kepler, we fitted the relationship of activity levels and $\mathrm{log} R^{\prime}_{HK}$ to ascertain their activity levels. Figure \ref{logR2level-p2p} depicts relationship between $\mathrm{log} R^{\prime}_{HK}$ and activity level (right panel). We found there is a quadratic relation between activity level and $\mathrm{log} R^{\prime}_{HK}$ in our samples and the Sun, {and the relations for stars of different spectral types (FGK) are consistent}. {Notably, the star with the highest activity level is HD 17925, an RS CVn Variable star considered an outlier and excluded during the fitting process.} We infer that this relation is applicable to other solar-type stars as well.

\subsection{Simulated spots and faculae}\label{sandf}
The primary parameter categories of spots and faculae include distribution, variability, contrast and size. We utilize the Python package {\tt\string butterpy} to generate spots with spatial and temporal distribution \citep{2022Claytor}.

Over the approximately 11-year solar activity cycle, spots emerge within active regions at high latitudes initially. Subsequently, these regions migrates toward the equator, indicating some overlap between consecutive cycles, resulting in a butterfly diagram illustrating the latitude distribution of spots over time. In our model, we set the overlap duration to be 10\% of cycle period. The butterfly diagram has also been observed in other stars \citep{2018Bazot, 2019Nielsen}. In our model, we apply the butterfly spot emergence feature of $\mathtt{butterpy}$. Because the faculae typically existed around spots, we add a facula near each spot in our model \citep{2015Borgniet}. The input parameters needed for $\mathtt{butterpy}$ are given in Table \ref{tab-input}. {Active regions are positioned at medium latitudes in our samples. Since polar spots have been observed in several fast rotators using Doppler and Zeeman Doppler imaging \citep{1983Vogt, 2009Strassmeier}, we also generated spots and faculae at latitudes ranging from $55^{\circ}$ to $85^{\circ}$ for samples with periods shorter than five days. We found that the difference in standard deviations of astrometric jitters between these two models (i.e., medium latitudes and high latitudes) is approximately 0.1 $\mu\mathrm{as}$. The latitudes of the activity regions are considered less influential due to the symmetry of our model.}

We estimated the temperature difference between the photosphere $T_{*}$ and spots $T_{\mathrm{sp}}$ {of FGK stars} \citep{2005Berdyugina, 2019Namekata}, revealing a trend with effective temperature $T_*$ of the star:
\begin{eqnarray}\label{csp}
\Delta T &=& T_{*} - T_{\mathrm{sp}}  \nonumber \\
&=& 3.58 \times 10^{-5} \, T_{*}  + 0.249 \, T_{*} - 808 \quad (\mathrm{K}) .
\end{eqnarray}

As the {area} ratio between penumbral and umbral of sunspots is approximately 4 to 5 \citep{2003Solanki}, the penumbral is significantly hotter than the umbra ($\sim 1000 \mathrm{K}$) \citep{2021Johnson}. Therefore, we weigh the differential temperature based on the relative areas in the penumbrae and umbrae within the range 0.2 and 0.6. The coefficient used is similar to that employed by \citet{2010Lagrange}.
The contrasts of stellar faculae are poorly constrained, we use an intensity contrast in the optical band {for G-type stars} \citep{2015Borgniet}:
\begin{eqnarray}
    C_{\textrm{fa}} = 0.131618 - 0.218744\mu + 0.104757\mu^2 ,
\end{eqnarray}
{the simulated spectra show that the differences in facular contrasts between K-type and G-type stars are relatively small at the band of CHES, and the contrasts of F-type stars are slightly lower than GK stars \citep{2019Norris, 2023Norris}.}
The contrasts of faculae are much lower than those of spots and are observed generated near spots. However their area is greater as observed on the Sun, the contribution from faculae is therefore expected to be significant. In this paper we use the area ratio to describe the size of faculae because we add a facula near each spot \citep{2015Borgniet}, the overlap situation was not considered in our simulations, given that the areas of features are significantly smaller compared to the stellar disk. We employed a lognormal distribution for spots in our simulations \citep{1988Bogdan, 2004Solanki, 2005Baumann}. Additionally, we used a lognormal distribution to describe the ratio between faculae and spots \citep{2015Borgniet}.

\begin{table*}
   \begin{center}
   \caption{Input parameters for spots and faculae} \label{tab-input}

    \begin{tabular}{clclc}
   \hline\noalign{\smallskip}
   \hline\noalign{\smallskip}
      Category  & Parameter &  Value/range & Unit& Reference \\
   \hline
      Stellar activity &  Cycle Period      & 0-60   &  yr  &     \\
                     &  Cycle Amplitude   & 0.1-5       &  -   &     \\
                     &  Cycle Overlap     & 0.1    &  -   &  \citet{2015Aigrain}  \\
   \hline
      Spots properties  & Mean initial size        & 46.5    & $\textrm{MSH}^{a}$    &  \\
                        & Standard size deviation  & 2.14    & MSH    &  \\
                        & Max size                 & 1500    & MSH    & \citet{2019Meunier}  \\
                        & Decay rate               &  43.9   & MSH/d   &  \\
                        & Max ave latitude         & 35  & deg &  \\
                        & Min ave latitude         &  7  & deg & \citet{2022Claytor}  \\
   \hline
      Faculae properties   & log Mean size ratio      & 0.8 & - & \\
                           & log Standard size ratio  & 0.4 & - &  \\
                           & Latitude deviation       &  3  & deg  &  \\
                           & Longitude deviation      &  3  & deg  &  \\
                           & Decay rate               & 21.9& MSH/d&  \citet{2015Borgniet} \\
   \hline
   \end{tabular}
   \end{center}
   \tablecomments{a:micro stellar hemispheres}
\end{table*}

\begin{figure}
   \centering
   \includegraphics[scale=0.50]{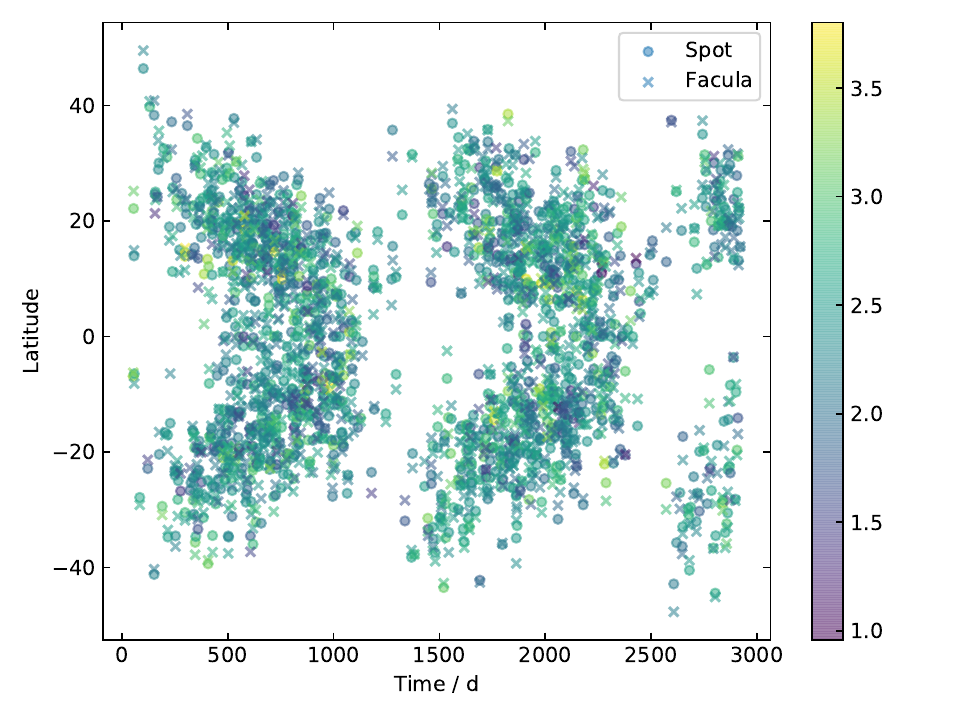}
   \caption{An example of the temporal and spatial evolution of spots and faculae on the surface of target 107 Psc, with a cycle period about 1200 days. Stellar magnetic features latitude distribution follows the butterfly diagram during eight years, the circles represent spot, and the crosses represent facula. The colorbar shows logarithmic areas of features, in unit of MSH (millionths of solar hemisphere).}
   \label{features distribution}
\end{figure}

Given that the emergence of spots and facula occurs within a few hours \citep{1992Howard}, {smaller than our time step, and we focus here on planets with periods of a few hundred days}, we neglected the growing process and assumed that all features are created with maximum area \citep{2015Borgniet}. We employed a parabolic decay law for both spots and faculae \citep{2020Gilbertson}.

All input parameters of spots and faculae for our model are summarized in Table \ref{tab-input}. For each star, the simulations last for eight years with a 0.9 day time step (total 3000 points) to ensure the simulation covers the entire cycle. At the beginning of each simulation no features are present on the stellar surface. Thus, we extracted the middle five years, assuming the simulation has reached a steady state. Figure \ref{features distribution} displays an example pattern of 107 Psc to show stellar magnetic features latitude distribution over eight years.

\section{Detection Efficiencies}
\label{sect:detectability}
Moreover, we used the inject-recover method to estimate detection limits of each star. We randomly generated 5000 planets in the parameters space. Since the main goal of CHES is to search Earth-like planets, we set the planetary mass range to of 0.5 to 5 Earth masses. All planets were placed in the habitable zone, with the inner and outer semi-major axis boundaries calculated based on the Runaway Greenhouse limit and Maximum Greenhouse limit, respectively \citep{2014Kopparapu}. All orbital elements and planetary masses followed a uniform distribution. The stellar wobbles due to a planet represent Keplerian orbit motion and can be derived through the Thiele-Innes equations \citep{1883Thiele}. Apart from the photocenter shifts due to stellar activity, we also account for other measurement errors, including photon noise, telescope error, and calibration errors. Gaussian noise, with a mean of 0 and standard deviations of 0.7 $\mu\mathrm{as}$ is added in the simulations \citep{2022Ji} .

For planets retrieval, the Lomb-Scargle periodograms are utilized \citep{1986Horne}. The period with the highest power is considered as a prior during fitting. Then we use a sinusoidal function based on the Thiele-Innes equations to fit planetary signal \citep{2022Jin}. In the fitting process, we assume all planets are in circular orbits to reduce computation. The amplitude of the sinusoidal function is denoted as $\alpha$, and the planetary mass is derived using Equation \ref{alpha}. We consider a planet to be recovered if both the derived period and mass within 30\% error is recovered. Additionally, we use the SNR criteria as a comparison. The SNR in astrometric planets is defined as \citep{2008Unwin}:
\begin{eqnarray}
   \mathrm{SNR} = \frac{\alpha}{\sigma} N_{vis}^{1/2} ,
\end{eqnarray}
where the $\sigma$ is the standard deviations of total noise, $N_{vis}$ is the number of observation, depending on observational strategy of CHES (Tan et al., in prep.). Table \ref{tab-tar} further presents \textrm{DHZ} (the distance to the center of habitable zone) for each target.

\begin{figure}
   \centering
   \includegraphics[scale=0.55]{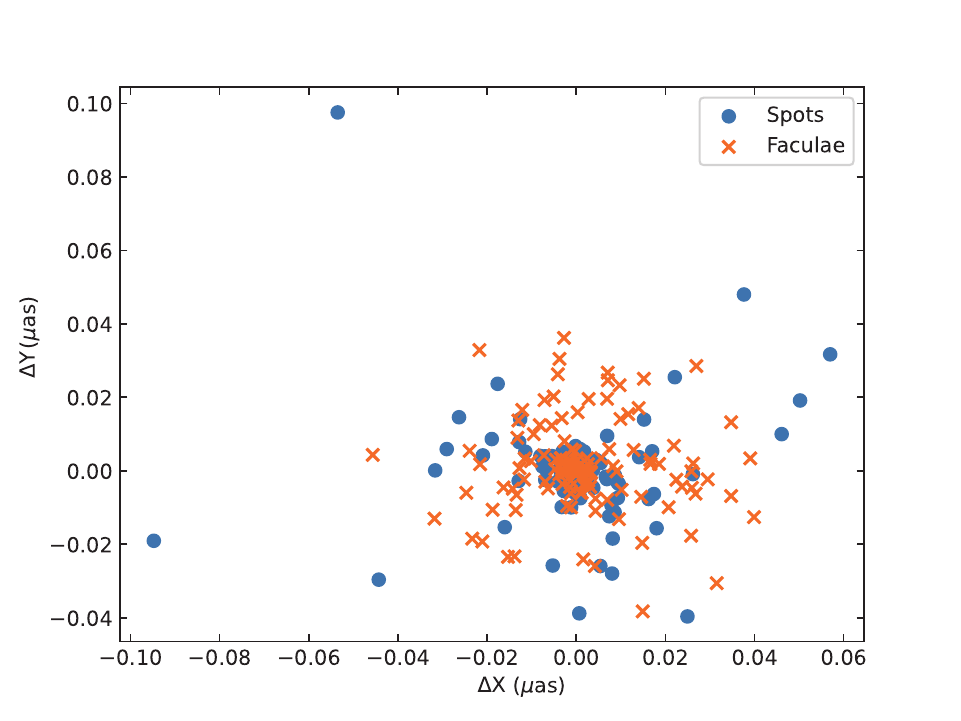}
   \caption{The photocenter jitter due to spots and faculae of HD 103095 during five years, the blue circles and orange crosses represent effects of spots and faculae, respectively.}
   \label{example_time}
\end{figure}

\section{Results of jitter amplitude and detection efficiency}
\label{sect:result}
\begin{figure}
   \centering
   \includegraphics[scale=0.5]{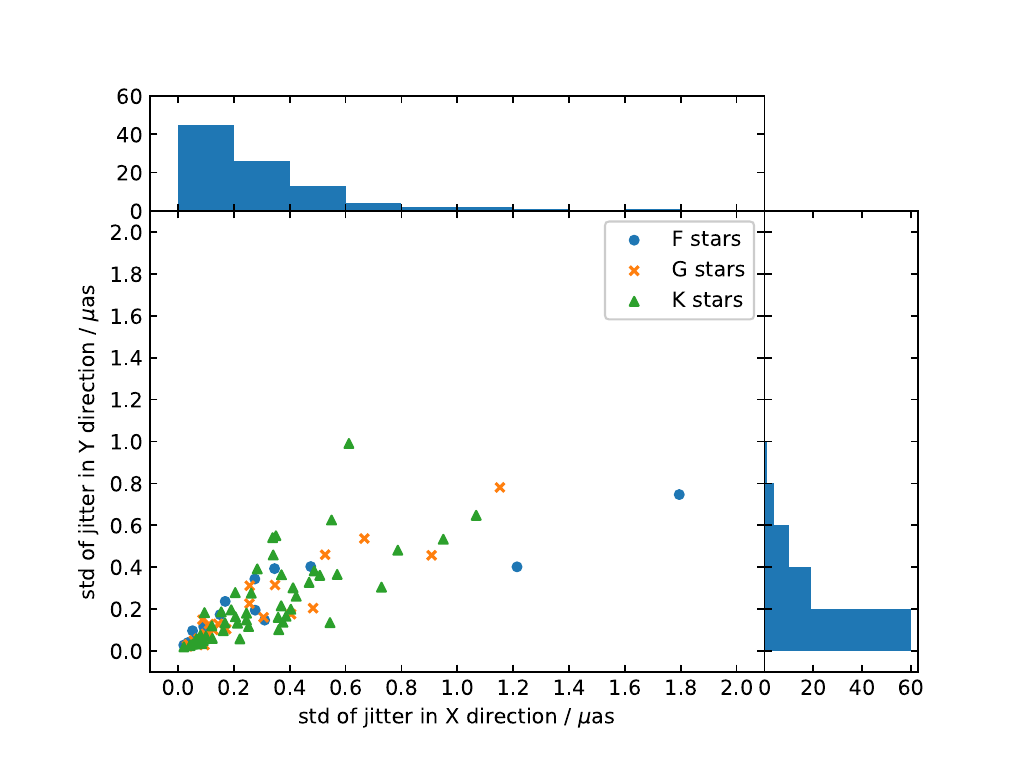}
   \caption{The standard deviations of astrometric jitter in two directions of all targets from simulations. The blue circle, orange cross, and green triangle correspond to F,G,K stars respectively. The two histograms shows jitter's distribution in our samples. }
   \label{hist1}
\end{figure}

\begin{figure}
   \centering
   \includegraphics[scale=0.5]{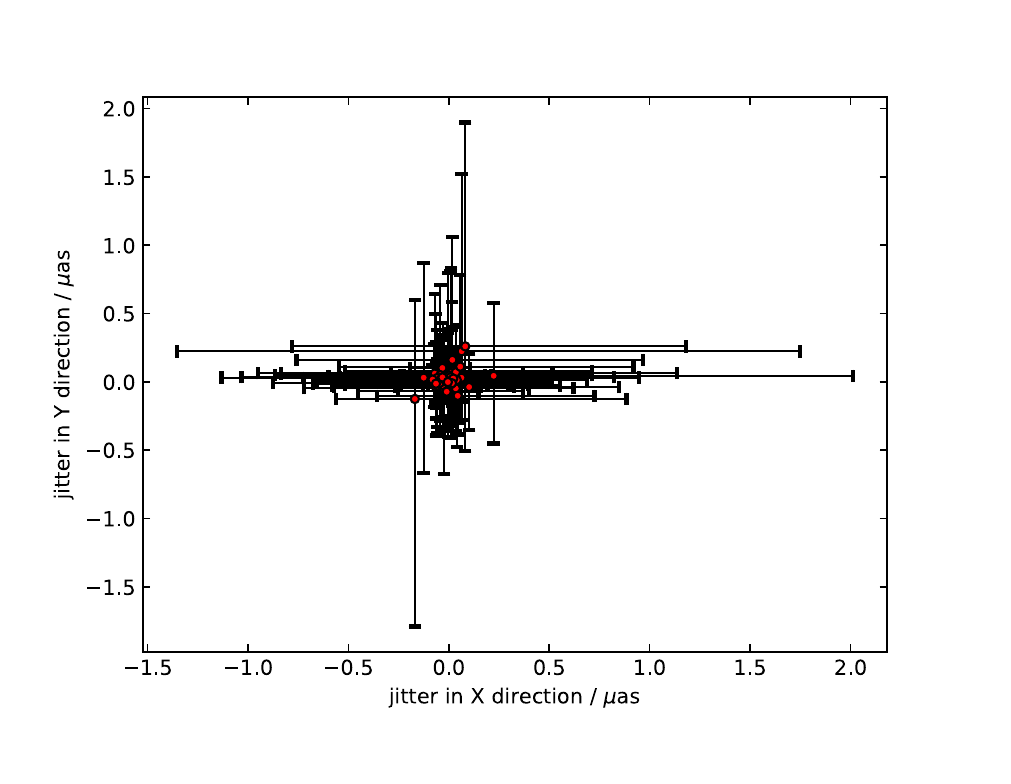}
   \caption{The mean jitter caused by stellar spots and faculae from simulations. The errorbars indicate the maximum and minimum jitter duration five years. }
   \label{jitter2}
\end{figure}

\begin{figure*}
   \gridline{\fig{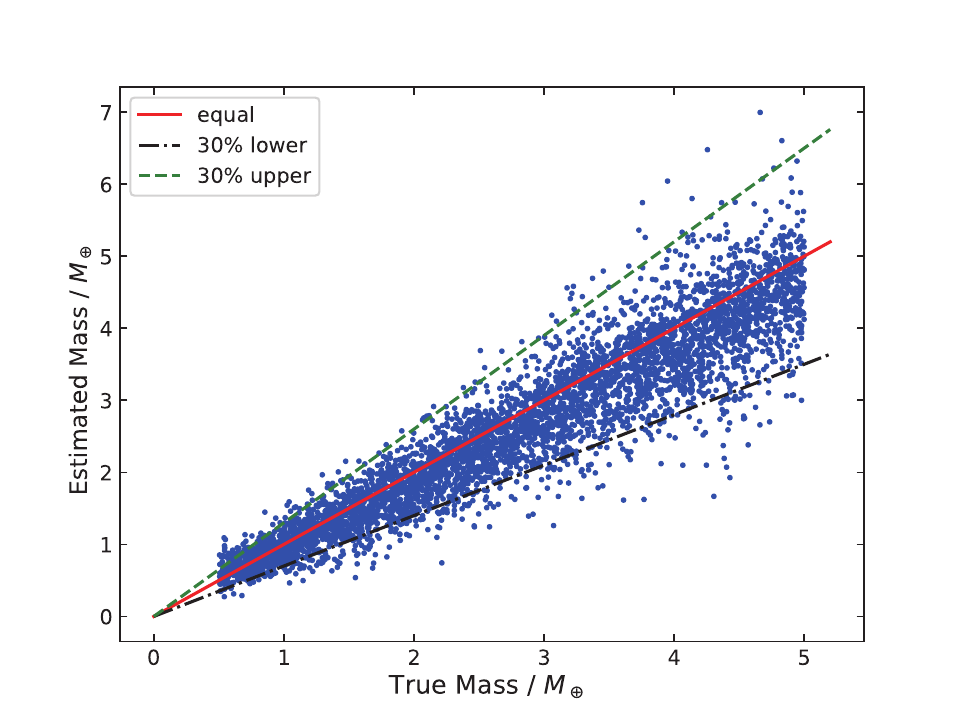}{0.45\textwidth}{(a)}
            \fig{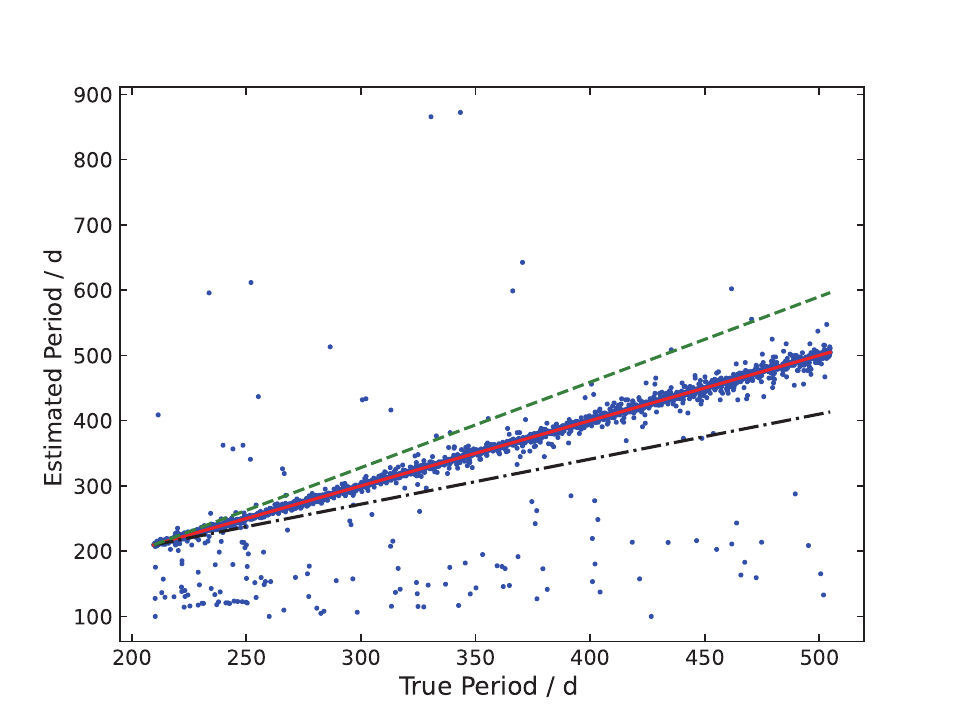}{0.45\textwidth}{(b)}}
   \caption{Properties of fitted mass (\emph{left panel}) and periods (\emph{right panel}) vs. true values of 107 Psc, every points indicates a injected planet. The red solid line indicates equal values, and the green dashed line and the black dashdotted indicates a 30\% upper and lower uncertainty level, respectively.
   \label{recover-mass-per}}
\end{figure*}

We applied the methods outlined in Section \ref{sect:method} to 94 stars. First we found there is a quadratic relation between activity level and the photometric peak-to-peak variations, as expected by \citet{2015Aigrain}. We then fitted a parabolic relation to calculate activity levels from $\mathrm{log} R^{\prime}_{HK}$ for our samples, and as the Sun corresponds to this relation as well, we infer that the relation can be extended to other FGK stars. Subsequently, we generated astrometric timeseries due to spots and faculae for each target. As an example, we consider the target HD 103095, a K-type star with no detected planets. The astrometric jitters of its photocenter are shown in Figure \ref{example_time}, with an amplitude about 0.1 $\mu \mathrm{as}$, which is lower than signal from an Earth-mass planet ($\sim$ 0.3 $\mu \mathrm{as} $). We derived the standard derivations in two directions of all targets, as shown in Figure \ref{hist1}. It was found that for more than 95\% of the targets, the standard deviations of photocenter shifts due to activity are less than 1 $\mu \mathrm{as}$. {Figure \ref{jitter2} shows the mean shifts of photocenter in our samples, which are within 0.4 $\mu \mathrm{as}$ for all stars. And the errorbars represents amplitudes of jitters. We found several stars with amplitudes larger than 1 $\mu \mathrm{as}$, because of shorter distance (e.g. Alpha Cen A and eps Eri) or the combination of higher activity level, larger size and short distance (e.g. eta Cas A).} We found the jitter in the Y direction is slightly smaller than that in the X direction because of the symmetry of butterfly diagram. And we derived the detection rates corresponding to terrestrial planets in habitable zone. The results for all targets are provided in Table \ref{tab-tar}. The inject-recover results of 107 Psc is shown in Figure \ref{recover-mass-per}. {We found about 95\% planets with $\mathrm{SNR}>6$ were recovered, consistent with those of \citet{2014Perryman}.}

\begin{figure*}
   \gridline{\fig{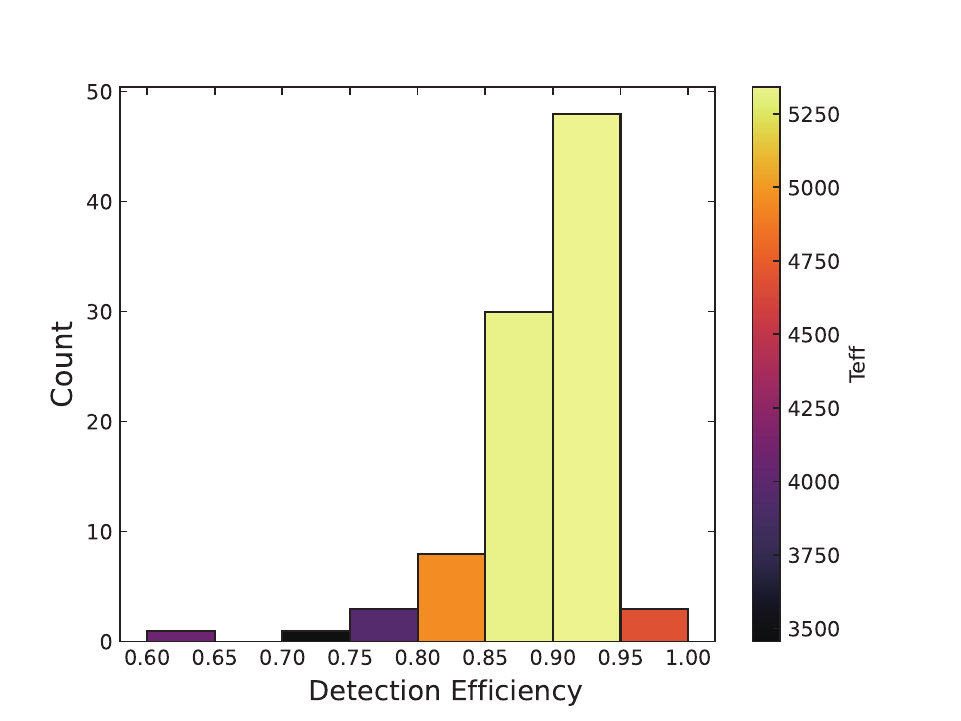}{0.45\textwidth}{(a)}
            \fig{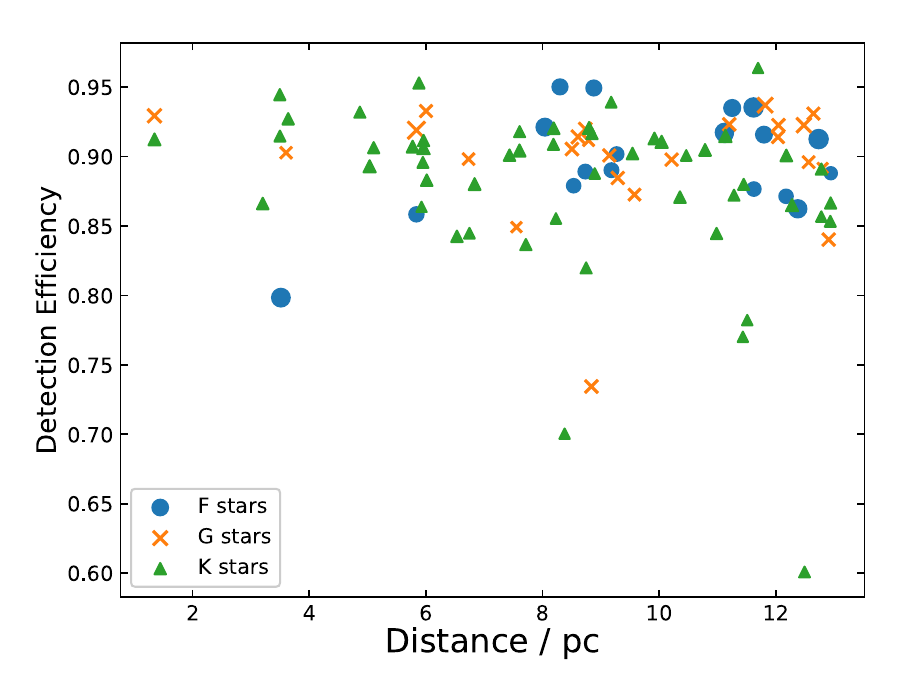}{0.45\textwidth}{(b)}}
   \caption{\emph{Left}: The histogram of detection efficiency, the colorbar show mean effective temperature in each bin. \emph{Right}: The detection efficiency and distance of all targets, sizes of each point indicate stellar radius. The blue circle, orange cross, and green triangle correspond to F,G,K stars respectively.
   \label{detect}}
\end{figure*}

The histogram of detection efficiency is shown in the left panel of Figure \ref{detect}, with nearly 95\% stars having a detection rate above 80\%. The detection efficiency decreases for some cold stars, such as HD 32450 and HD 21531, due to the weak planetary signal from a closer habitable zone. Despite several close stars, such as $\alpha$ Cen A and eps Eri having an obvious jitter, their detection efficiencies remain high because the planetary signals also increase as the distance gets closer. {The right panel of Figure \ref{detect} displays the detection efficiencies versus the distance of the stars. In our sample, six stars have detection efficiencies below 80\%; five of them are located 8 \textrm{pc} away, exhibiting a slightly weak planetary signal. Notably, the subgiant alf CMi has a detection efficiency of about 79.8\% due to its much more distant habitable zone. The period corresponding to habitable planets around alf CMi is approximately the same length as the time baseline of CHES, making planet characterization difficult.}

\section{Conclusions and Discussions}
\label{sect:conclusion}

In this study, we conducted simulations to evaluate the influence of stellar activity on the detection of terrestrial planets in the context of the ultra-high precision astrometry space mission, CHES, which aims to perform measurements on approximately 100 nearby stars. Our model, implemented within $\mathtt{butterpy}$, generated simulated astrometry data for the identified CHES targets. {We estimated the contrast of spots based on Equation \ref{csp}, which applies to stars with effective temperatures between 3000 to 7000 \textrm{K} \citep{2005Berdyugina, 2019Namekata}. Notably, there are about a dozen stars beyond this temperature range. As the empirical formula is believed to overestimate contrasts, especially for hotter stars \citep{2015Borgniet,2019Meunier}. \citet{2015Borgniet} also estimated a lower bound of spots contrasts. In this paper, we choose Equation \ref{csp} as an extreme case because higher contrasts cause larger astrometric noise. Similarly, the facular effect of F-type stars can also be overestimated in our simulation as introduced in Section \ref{sandf}. Therefore, higher detection efficiency of CHES is actually possible.}

{As some peak-to-peak values were derived from multiple sectors, there are errors in each star's activity level, as depicted in Figure \ref{logR2level-p2p}. The astrometric jitters of ten targets with larger errorbars were calculated based on their maximum activity level value. Tet Per exhibits a noise about 1.3 $\mu \mathrm{as}$ with an activity level of approximately 2.5. While noise from stellar activity is still not the dominant factor for other stars, with standard deviations lower than 0.7 $\mu \mathrm{as}$. These ten stars are indicated in Table \ref{tab-tar}. }

The determination of each star's activity level relied on the chromospheric index $\mathrm{log} R^{\prime}_{HK}$. Our observations revealed a discernible correlation between increased activity levels and heightened variability in both photometry and astrometry, best described by a parabolic relation. Subsequent analysis of the time series facilitated the estimation of detection efficiencies for planets within the habitable zone of each target. Notably, the influence of stellar activity on astrometry was found to be minimal, with photocenter jitter measuring less than 1 $\mu \mathrm{as}$ for around 95\% of the targets. In the context of terrestrial planet detection, the astrometric method demonstrated superior efficacy compared to current radial velocity and transit methods.

{We compared our findings with the results of \citet{2022Meunier}, who estimated the astrometric noise from stellar activity of Theia's targets. Among the 37 common targets, the differences in astrometric jitter are less than 0.3 $\mu \mathrm{as}$ for approximately 80\% of the stars. }

In this work, we intentionally exclude the consideration of other stellar characteristics, such as oscillation, granulation, and supergranulation, as their expected impact on astrometry is deemed negligible \citep{2021Sowmya}. The associated timescales \citep{1999Staude, 2015Borgniet, 2017Chiavassa} are considerably shorter than the observation cadence of the Closeby Habitable Exoplanet Survey \citep{2022Ji}. Our focus remains on single-planet systems, although multi-planet systems are prevalent in observational datasets. Approximately 10 targets in our sample are identified as hosting hot Jupiters or Neptunes based on radial velocity measurements. The presence of existing gas-giants may obscure signals from Earth-like planets, a factor we plan to address in forthcoming study.

In our simulations, stellar inclinations are consistently set in an edge-on configuration. While \citet{2019Meunier} concluded that the amplitude of radial velocity reaches a maximum in this configuration, \citet{2021Sowmya} found that the largest astrometric jitter occurs at intermediate inclinations. A nuanced exploration of the impact of inclinations is scheduled for future investigations. {In this work, we do not consider the growth phase of activity regions, which is not well understood observationally at present. Since the growth rates of sunspots are much faster than their decay rates \citep{1992Howard, 2021Forg}, and nearly 90\% of the targets in our sample exhibit higher activity levels than the Sun, we chose to disregard this effect in our study. The total flux contribution, including the growth process, is estimated to be less than 1.0\% based on simulations by \citet{2015Aigrain}. Assuming that the dip in flux originates from only one spot on the edge, the photocenter shifts are approximately 0.4 $\mu \mathrm{as}$ for a solar-like star at a distance of 10 \textrm{pc}, which is comparable to the results obtained from our no-growth model. Given the limited understanding of spot evolution, this growth phase will be incorporated in future study to provide a more comprehensive description of the astrometric noise from stellar activity at small timescales.} This approach aims to contribute to a more holistic understanding of the challenges and opportunities associated with astrometric detection methods, especially when considering diverse planetary systems. Furthermore, our forthcoming research will systematically investigate additional factors influencing the astrometric precision of CHES, including comprehensive considerations such as the potential presence of planets around the reference stars and the stability of the satellite.

\section*{acknowledgements}
\begin{acknowledgements}
   We thank the referee for constructive comments and suggestions to improve the manuscript. This work is financially supported by the National Natural Science Foundation of China (grant Nos. 12033010 and 11773081), the Strategic Priority Research Program on Space Science of the Chinese Academy of Sciences (Grant No. XDA 15020800), and the Foundation of Minor Planets of the Purple Mountain Observatory.
\end{acknowledgements}

\software{astropy \citep{2013Astropy},
butterpy \citep{2022Claytor},
lightkurve \citep{2018Lightkurve},
matplotlib \citep{2007Hunter},
PyMsOfa \citep{2023Ji},
scipy \citep{2020Virtanen}}
\hfill

\appendix

\begin{longrotatetable}
\begin{deluxetable*}{cchccccccccccccc}
   \tablecaption{The parameters of target stars for CHES mission}\label{tab-tar}
   \tablenum{2}
   \tablehead{\colhead{identifier} & \colhead{Level} & & \colhead{$T_{eff}$} & \colhead{log$g$} & \colhead{radius} & \colhead{mass} & \colhead{$D$} & \colhead{TS} & \colhead{ $\mathrm{log} R^\prime_{\mathrm{HK}}$} & \colhead{$P_{rot}$} & \colhead{$P_{cyc}$} & \colhead{DHZ} & \colhead{$\sigma(\Delta x$)} & \colhead{$\sigma(\Delta y$)} & \colhead{efficiency} \\
   \colhead{} & \colhead{} & \colhead{} & \colhead{(K)} & \colhead{} & \colhead{($R_{\odot}$)} & \colhead{($M_{\odot}$)} & \colhead{(pc)} & \colhead{} & \colhead{} & \colhead{(d)} & \colhead{(d)} & \colhead{(au)} & \colhead{($\mu \mathrm{as}$)} & \colhead{($\mu \mathrm{as}$)} & \colhead{} }

   \startdata
   107 Psc  & 1.03 & 65 & 5190 & 4.54 & 0.83 & 0.88 & 7.61 & K1V & -5.00(23) & 51.6 & 1241.1 & 0.93 & 0.059 & 0.029 & 0.904 \\
   11 Lmi(*)   & 2.13 & 109 & 5499 & 4.43 & 0.98 & 0.96 & 11.20 & G8V & -4.66(23)& 22.4 & 1144.1 & 1.20 & 0.22 & 0.158 & 0.923 \\
   12 Oph   & 1.78 & 117 & 5293 & 4.58 & 0.81 & 0.91 & 9.92 & K0V & -4.72(15)& 25.9 & 1161.0 & 0.93 & 0.089 & 0.127 & 0.913 \\
   36 Oph A & 1.80 & 71 & 5121 & 4.52(6) & 0.89(6) & 0.96(6)& 5.96 & K2V  & -4.74(23)& 28.7 & 1302.8 & 0.80 & 0.385 & 0.18 & 0.906 \\
   36 Oph B & 1.95 & 71 & 5143 & 4.66 & 0.72 & 0.87 & 5.96 & K1V & -4.66(3) & 25.8 & 1311.2 & 0.79 & 0.288 & 0.16 & 0.912 \\
   41 Ara A(*) & 1.16 & 83 & 5380 & 4.63 & 0.77 & 0.93 & 8.79 & G9V & -4.94(23) & 48.4 & 1188.3 & 0.91 & 0.07 & 0.057 & 0.912 \\
   54 Psc & 1.17 & 138 & 5275 & 4.51 & 0.87 & 0.90 & 11.14 & K0.5V & -5.05(15) & 49.5 & 1260.8 & 1.00 & 0.058 & 0.044 & 0.915 \\
   61 Cyg A & 1.42 & 78 & 4305 & 4.56 & 0.71 & 0.67 & 3.49 & K5V & -4.91(15)& 47.9 & 1489.6 & 0.58 & 0.069 & 0.161 & 0.945 \\
   61 Cyg B & 1.86 & 78 & 3948 & 4.55 & 0.69 & 0.61 & 3.50 & K7V & -4.96(15)& 54.4 & 1571.7 & 0.48 & 0.327 & 0.241 & 0.915 \\
   61 UMa   & 1.99 & 88 & 5502 & 4.55 & 0.87 & 0.97 & 9.58 & G8V & -4.53(15)& 14.0 & 1084.5 & 1.06 & 0.101 & 0.082 & 0.873 \\
   61 Vir   & 0.87 & 58 & 5562 & 4.44 & 0.98 & 0.98 & 8.51 & G7V & -5.01(15)& 33.6 & 897.2 & 1.22 & 0.05 & 0.045 & 0.905 \\
   70 Oph B & 1.65 & 30 & 4360(27) & 4.57(27) & 0.71(27)& 0.70(27)& 5.11 & K4V & -4.76(3)& 37.1 & 1517.8 & 0.75 & 0.364 & 0.215 & 0.906 \\
   85 Peg   & 0.91 & 124 & 5336 & 4.45(34)& 0.91(37)& 0.89(37)& 12.65 & G5V& -4.84(23)& 23.2 & 807.8 & 1.28 & 0.041 & 0.018 & 0.931 \\
   alf Cen A & 1.19 & 30 & 5585(11)& 4.31(31) & 1.11(30)& 1.13(30)& 1.35 & G2V  & -5.01(9)& 34.9 & 926.2 & 1.64 & 0.332 & 0.202 & 0.929 \\
   alf Cen B& 1.31 & 30 & 5151(30) & 4.38(30) & 0.81(30) & 0.86(30) & 1.35 & K1V & -4.92(5) & 43.7 & 1335.9 & 0.94 & 0.048 & 0.037 & 0.912 \\
   alf CMi  & 1.50 & 30 & 6775(30) & 3.74(30) & 1.82(30) & 1.53(30) & 3.51 & F5IV & -4.82(3) & 2.9 & 106.0 & 3.22 & 0.056 & 0.033 & 0.798 \\
   alf Men & 1.35 & 81 & 5590(30) & 4.20(30) & 0.95(30) & 0.98(30) & 10.21 & G7V & -4.89(15) & 29.8 & 951.79 & 1.24 & 0.015 & 0.017 & 0.898 \\
   bet Com & 1.45 & 60 & 5969 & 4.39 & 1.11 & 1.10 & 9.18 & F9.5V & -4.77(30) & 13.3 & 525.3 & 1.54 & 0.124 & 0.088 & 0.890 \\
   bet CVn & 0.93 & 66 & 5884 & 4.40 & 1.08 & 1.07 & 8.61 & G0V & -4.98(30) & 21.1 & 586.4 & 1.46 & 0.143 & 0.155 & 0.914 \\
   bet Hyi & 1.20 & 31 & 5901(30) & 4.14(19) & 1.76(19) & 1.12(19) & 5.84 & G0V & -5.05(30) & 24.4 & 618.46 & 2.15 & 0.086 & 0.106 & 0.919 \\
   bet TrA & 1.55 & 121 & 7108 & 4.22(6) & 1.73(27) & 1.58(27) & 12.38 & F1V & -4.80(1) & 5.3 & 199.3 & 1.60 & 0.034 & 0.024 & 0.862 \\
   bet Vir & 1.24 & 47 & 6071 & 4.00 & 1.75 & 1.13 & 11.12 & F9V & -4.96(15) & 13.7 & 390.0 & 2.49 & 0.191 & 0.188 & 0.917 \\
   chi Dra & 1.09 & 56 & 6119 & 4.28(16) & 1.31 & 1.20(27) & 8.30 & F7V & -4.80(1) & 15.4 & 581.4 & 1.89 & 0.114 & 0.114 & 0.950 \\
   chi01 Ori & 3.15 & 60 & 5988 & 4.49 & 0.99 & 1.10 & 8.84 & G0V & -4.38(23) & 3.5 & 587.1 & 1.38 & 0.285 & 0.157 & 0.734 \\
   e Eri & 0.84 & 49 & 5413 & 4.48 & 0.92 & 0.94 & 6.00 & G6V & -5.00(15) & 34.1 & 928.6 & 1.09 & 0.306 & 0.27 & 0.933 \\
   eps Eri & 2.2 & 49 & 5065 & 4.61 & 0.76 & 0.85 & 3.20 & K2V & -4.6(15) & 22.4 & 1366.8 & 0.81 & 1.223 & 1.305 & 0.866 \\
   eps Ind & 1.58 & 31 & 4611 & 4.57 & 0.74 & 0.73 & 3.64 & K5V & -4.85(15) & 42.0 & 1441.1 & 0.67 & 0.645 & 0.147 & 0.927 \\
   eta Cas A(*) & 2.32 & 30 & 5919 & 4.41(34) & 1.19(37) & 1.03(37) & 5.84 & F9V & -4.99(3) & 17.5 & 484.6 & 1.46 & 0.251 & 0.212 & 0.864 \\
   eta Cas B & 2.12 & 30 & 3973(37) & 4.56(37) & 0.60(37) & 0.60(37) & 5.93 & K7Ve & -4.94(3) & 53.3 & 1590.2 & 0.42 & 0.069 & 0.033 & 0.858 \\
   gam Lep & 1.17 & 113 & 6258 & 4.29 & 1.30 & 1.22 & 8.88 & F6V & -4.97(15) & 6.6 & 187.2 & 1.94 & 0.44 & 0.235 & 0.949 \\
   gam Pav & 0.85 & 60 & 6109 & 4.43 & 1.08 & 1.15 & 9.27 & F9V & -4.81(15) & 5.7 & 210.9 & 1.55 & 0.273 & 0.163 & 0.902 \\
   gam Ser & 0.78 & 59 & 6236 & 4.18 & 1.48 & 1.21 & 11.25 & F6IV & -4.74(23) & 6.1 & 256.0 & 2.19 & 0.109 & 0.071 & 0.935 \\
   gam Vir A & 0.80 & 116 & 7147 & 4.04(24) & 1.95(6) & 1.39(35) & 11.62 & F1-F2V & -5.04(15) & 12.2 & 313.1 & 2.56 & 0.056 & 0.024 & 0.935 \\
   gam Vir B & 3.74 & 116 & 7090(8) & 4.49(8) & 1.95(35) & 1.37(35) & 12.74 & F0mF2V & -4.30(23) & 1.5 & 355.7 & 2.30 & 0.227 & 0.341 & 0.913 \\
   i Boo A & 2.22 & 122 & 5820(33) & 4.33(32) & 0.87(33) & 0.97(33) & 12.94 & F5V & -4.59(23) & 12.3 & 761.3 & 1.33 & 0.106 & 0.063 & 0.888 \\
   i Boo B & 1.55 & 122 & 5820(33) & 4.11(6) & 0.66(33) & 0.55(33) & 12.81 & G9 & -4.80(10) & 24.7 & 930.2 & 1.33 & 1.04 & 0.87 & 0.892 \\
   iot Peg & 1.67 & 51 & 6565 & 4.25 & 1.45 & 1.36 & 11.80 & F5V & -4.67(23) & 2.4 & 122.3 & 2.32 & 0.117 & 0.05 & 0.916 \\
   kap01 Cet & 2.93 & 70 & 5712 & 4.50 & 0.95 & 1.02 & 9.15 & G5V & -4.39(15) & 7.6 & 835.4 & 1.22 & 0.049 & 0.026 & 0.901 \\
   ksi Boo A & 2.72 & 50 & 5496 & 4.53 & 0.86 & 0.90 & 6.73 & G7Ve & -4.44(3) & 9.4 & 1076.5 & 1.00 & 0.087 & 0.077 & 0.898 \\
   ksi Boo B & 3.04 & 50 & 4288 & 4.64 & 0.65 & 0.67 & 6.75 & K5Ve & -4.42(1) & 12.3 & 1610.8 & 0.53 & 0.382 & 0.33 & 0.845 \\
   ksi UMa A & 3.85 & 82 & 5977(34) & 4.19(23) & 1.10 & 1.15(21) & 8.73 & F8.5 & -4.28(23) & 1.7 & 446.3 & 1.68 & 0.028 & 0.016 & 0.889 \\
   ksi UMa B & 3.87 & 82 & 5476(34) & 5.04(34) & 1.00(37) & 0.96(29) & 8.73 & G2V & -4.29(23) & 3.3 & 821.8 & 1.47 & 0.054 & 0.045 & 0.923 \\
   lam Aur & 0.91 & 70 & 5873 & 4.25 & 1.28 & 1.06 & 12.48 & G1.5IV & -5.09(34) & 26.9 & 651.8 & 1.73 & 0.163 & 0.215 & 0.923 \\
   lam Ser & 0.84 & 119 & 5901 & 4.20 & 1.35 & 1.07 & 11.82 & G0-V & -4.92(23) & 17.1 & 1506.3 & 1.84 & 0.051 & 0.049 & 0.937 \\
   mu. Cas & 2.15 & 30 & 5316 & 4.70(32) & 0.68 & 0.72(17) & 7.55 & G5Vb & -4.99(3) & 31.5 & 865.3 & 0.88 & 0.06 & 0.043 & 0.849 \\
   omi02 Eri & 1.33 & 30 & 5092 & 4.51 & 0.85 & 0.85 & 5.04 & K0V & -4.93(15) & 40.4 & 1221.9 & 0.91 & 0.354 & 0.43 & 0.893 \\
   p Eri A & 1.28 & 119 & 5020 & 4.58 & 0.78 & 0.83 & 8.19 & K2V & -4.90(9) & 40.8 & 1291.0 & 0.82 & 0.099 & 0.066 & 0.920 \\
   p Eri B & 2.66 & 119 & 5107 & 4.68 & 0.70 & 0.86 & 8.19 & K2V & -4.83(9) & 37.8 & 1347.5 & 0.76 & 0.337 & 0.237 & 0.909 \\
   pi.03 Ori(*) & 2.14 & 30 & 6398 & 4.05(18) & 1.65(37) & 1.25(37) & 8.04 & F6V & -4.65(7) & 2.6 & 138.3 & 2.12 & 0.13 & 0.151 & 0.921 \\
   sig Dra & 1.02 & 39 & 5242 & 4.59 & 0.80 & 0.90 & 5.77 & K0V & -4.92(15) & 37.2 & 1136.2 & 0.90 & 0.615 & 0.479 & 0.907 \\
   tau Cet & 1.21 & 30 & 5333 & 4.56 & 0.83 & 0.92 & 3.60 & G8V & -5.00(23) & 35.3 & 957.2 & 0.97 & 0.311 & 0.235 & 0.903 \\
   tet Per(*) & 1.65 & 67 & 6196 & 4.27 & 1.32 & 1.19 & 11.13 & F8V & -4.70(15) & 6.4 & 296.4 & 1.94 & 0.864 & 0.545 & 0.916 \\
   zet Dor & 3.52 & 85 & 6153 & 4.45 & 1.07 & 1.17 & 11.63 & F9VFe-0.5 & -4.39(15) & 2.2 & 367.9 & 1.55 & 0.355 & 0.342 & 0.877 \\
   zet TrA & 2.87 & 102 & 6078 & 4.45 & 1.05 & 1.14 & 12.18 & F9V & -4.50(23) & 5.1 & 431.1 & 1.50 & 0.085 & 0.069 & 0.871 \\
   zet Tuc & 0.70 & 51 & 5961 & 4.43 & 1.05 & 1.09 & 8.53 & F9.5V & -5.22(15) & 22.3 & 463.8 & 1.45 & 0.064 & 0.06 & 0.879 \\
   zet01 Ret & 2.16 & 139 & 5744 & 4.54 & 0.90 & 1.03 & 12.04 & G2.5 & -4.60(15) & 12.0 & 727.1 & 1.17 & 0.065 & 0.051 & 0.914 \\
   zet02 Ret & 1.08 & 139 & 5866 & 4.48 & 0.98 & 1.06 & 12.05 & G1V & -4.73(15) & 13.1 & 566.8 & 1.32 & 0.498 & 0.217 & 0.923 \\
   CD-57 1079 & 3.11 & 135 & 4186(18) & 4.66 & 0.59 & 0.58 & 11.70 & K7Vk & -4.47(23) & 17.0 & 1700.6 & 1.32 & 0.097 & 0.095 & 0.964 \\
   HD 100623 & 1.13 & 120 & 5140 & 4.61 & 0.77 & 0.87 & 9.54 & K0V & -4.91(15) & 38.9 & 1203.9 & 0.84 & 0.108 & 0.031 & 0.902 \\
   HD 101581 & 1.71 & 115 & 4634 & 4.71 & 0.63 & 0.74 & 12.78 & K4.5Vk & -4.74(15) & 34.6 & 1459.9 & 0.58 & 0.176 & 0.14 & 0.857 \\
   HD 102365 & 0.99 & 55 & 5672 & 4.49 & 0.95 & 1.01 & 9.29 & G2V & -4.94(15) & 19.9 & 584.5 & 1.22 & 0.638 & 0.552 & 0.885 \\
   HD 103095 & 1.41 & 141 & 4934 & 4.70 & 0.66 & 0.81 & 9.18 & K1V & -4.48(13) & 6.7 & 628.4 & 0.68 & 0.303 & 0.087 & 0.939 \\
   HD 10780 & 1.72 & 97 & 5344 & 4.54 & 0.85 & 0.92 & 10.04 & K0V & -4.70(23) & 50.0 & 1232.1 & 0.99 & 0.98 & 4.62 & 0.911 \\
   HD 131977 & 1.38 & 111 & 4681 & 4.60 & 0.72 & 0.75 & 5.88 & K4V  & -4.88(23) & 44.4 & 1461.1 & 0.67 & 0.421 & 0.244 & 0.953 \\
   HD 13445 & 1.37 & 136 & 5189 & 4.59 & 0.79 & 0.88 & 10.79 & K1.5V & -4.76(23) & 25.3 & 1135.7 & 0.88 & 0.361 & 0.81 & 0.905 \\
   HD 147513(*) & 3.01 & 117 & 5873 & 4.49 & 0.97 & 1.06 & 12.91 & G5V & -4.48(15) & 7.2 & 685.4 & 1.30 & 0.605 & 0.297 & 0.840 \\
   HD 154363 & 1.66 & 115 & 4360 & 4.64 & 0.65 & 0.68 & 10.47 & K4/5V & -4.76(23) & 36.0 & 1451.2 & 0.54 & 0.289 & 0.273 & 0.901 \\
   HD 156384 & 1.88 & 40 & 4819 & 4.50(2) & 0.80(36) & 1.26(12) & 6.84 & K3V+K5V & -4.69(23) & 30.7 & 1479.6 & 0.98 & 0.022 & 0.011 & 0.882 \\
   HD 157881 & 2.87 & 39 & 4035 & 4.53 & 0.72 & 0.63 & 7.71 & K7V & -4.45(23) & 15.3 & 1664.7 & 0.52 & 0.015 & 0.037 & 0.837 \\
   HD 17925(*) & 6.58 & 140 & 5225 & 4.61 & 0.77 & 0.89 & 10.36 & V-K1V & -4.32(15) & 1.6 & 453.1 & 0.87 & 0.069 & 0.055 & 0.871 \\
   HD 191408 & 1.67 & 30 & 5044 & 4.23 & 0.76 & 0.78 & 6.02 & K2.5V & -5.04(23) & 52.6 & 1285.5 & 0.75 & 0.312 & 0.206 & 0.883 \\
   HD 192310 & 0.90 & 107 & 5071 & 4.54 & 0.82 & 0.85 & 8.80 & K2+V & -5.12(23) & 55.5 & 1302.0 & 0.88 & 0.161 & 0.125 & 0.921 \\
   HD 21531 & 3.27 & 133 & 4231(22) & 4.67(22) & 0.67(25) &0.66(37) & 12.49 & K5V & -4.38(23) & 9.8 & 1728.5 & 0.43 & 0.13 & 0.082 & 0.601 \\
   HD 217357 & 2.83 & 140 & 4081 & 4.64 & 0.63 & 0.64 & 8.23 & K7+Vk & -4.50(23) & 19.5 & 1682.5 & 0.47 & 0.212 & 0.143 & 0.855 \\
   HD 219134 & 1.51 & 30 & 4884 & 4.59 & 0.75 & 0.80 & 6.53 & K3V & -4.92(23) & 46.1 & 1412.8 & 0.76 & 0.447 & 0.497 & 0.843 \\
   HD 222237 & 1.15 & 98 & 4752 & 4.62 & 0.71 & 0.76 & 11.45 & K3+V & -5.10(15) & 57.7 & 1369.7 & 0.68 & 0.722 & 1.017 & 0.880 \\
   HD 32147 & 0.97 & 121 & 4931 & 4.61 & 0.74 & 0.81 & 8.85 & K3+V & -4.97(23) & 50.2 & 1415.3 & 0.75 & 0.361 & 0.405 & 0.917 \\
   HD 32450 & 2.22 & 60 & 3455(26) & 4.35(26) & 0.62 & 0.61 & 8.38 & K7V & -4.62(26) & 29.4 & 1670.7 & 0.55 & 0.249 & 0.266 & 0.700 \\
   HD 36003 & 1.48 & 118 & 4616 & 4.62 & 0.69 & 0.73 & 12.94 & K5V & -4.95(15) & 49.9 & 1452.4 & 0.63 & 0.604 & 0.239 & 0.853 \\
   HD 37394(*) & 3.07 & 118 & 5249 & 4.55 & 0.84 & 0.90 & 12.28 & K1 & -4.56(15) & 18.9 & 1318.2 & 0.95 & 0.064 & 0.058 & 0.865 \\
   HD 38A(*) & 2.30 & 125 & 3959 & 4.65 & 0.62 & 0.62 & 11.51 & K6V & -4.49(23) & 19.1 & 1703.0 & 0.43 & 0.062 & 0.033 & 0.782 \\
   HD 40307 & 1.15 & 103 & 4867 & 4.63 & 0.72 & 0.79 & 12.94 & K2.5V & -5.16(15) & 60.7 & 1332.6 & 0.72 & 0.283 & 0.195 & 0.867 \\
   HD 4628 & 1.21 & 80 & 4937 & 4.60 & 0.75 & 0.81 & 7.44 & K2.5V & -4.98(15) & 46.7 & 1307.7 & 0.77 & 0.168 & 0.149 & 0.901 \\
   HD 50281 & 2.74 & 50 & 4719 & 4.63 & 0.70 & 0.76 & 8.75 & K3.5V & -4.55(15) & 21.4 & 1505.7 & 0.67 & 0.049 & 0.041 & 0.820 \\
   HD 69830(*) & 1.08 & 141 & 5387 & 4.50 & 0.89 & 0.93 & 12.56 & G8:V & -4.99(15) & 41.7 & 1153.3 & 1.06 & 0.326 & 0.191 & 0.896 \\
   HD 72673 & 1.30 & 140 & 5213 & 4.60 & 0.78 & 0.89 & 12.18 & K1V & -4.95(15) & 39.4 & 1157.3 & 0.87 & 0.428 & 0.189 & 0.901 \\
   HD 82106 & 3.16 & 140 & 4828 & 4.62 & 0.71 & 0.78 & 12.78 & K3V & -4.50(23) & 17.1 & 1490.9 & 0.70 & 0.063 & 0.049 & 0.891 \\
   HD 85512 & 1.39 & 92 & 4433 & 4.60 & 0.69 & 0.69 & 11.28 & K6Vk & -4.99(15) & 52.9 & 1470.9 & 0.59 & 0.235 & 0.13 & 0.872 \\
   HD 88230 & 1.24 & 125 & 3893(20) & 4.57 & 0.70 & 0.67 & 4.87 & K6V & -4.47(23) & 33.0 & 1603.4 & 0.51 & 0.164 & 0.081 & 0.932 \\
   LHS 2713 & 3.82 & 135 & 5019 & 4.63 & 0.73 & 0.83 & 10.99 & K2V & -4.50(15) & 17.1 & 1450.7 & 0.77 & 0.541 & 0.512 & 0.845 \\
   V AK Lep & 1.98 & 113 & 5250 & 4.78 & 0.64 & 0.90 & 8.90 & K3 & -4.70(15) & 27.3 & 1410.5 & 0.72 & 0.138 & 0.112 & 0.888 \\
   V TW PsA & 3.11 & 101 & 4617 & 4.62 & 0.69 & 0.73 & 7.61 & K4Ve & -4.31(15) & 6.5 & 1674.1 & 0.63 & 0.03 & 0.035 & 0.918 \\
   V V2215 Oph & 2.34 & 57 & 4385 & 4.60 & 0.69 & 0.69 & 5.95 & K5V & -4.56(15) & 22.7 & 1547.2 & 0.58 & 0.098 & 0.112 & 0.896 \\
   V V2689 Ori & 5.14 & 125 & 3936 & 4.64 & 0.62 & 0.60 & 11.44 & K6V & -4.14(23) & 4.5 & 1819.8 & 0.43 & 0.03 & 0.011 & 0.770 \\
\enddata


   \tablecomments{{TS presents stellar spectral types, which are obtained from CDS. Level is the stellar activity introduced in Section \ref{sect:level}. $T_{eff}$ is the effective temperature of star. log$g$ is the stellar surface gravity. $D$ is the distance of stars. $P_{rot}$ and $P_{cyc}$ present the stellar rotation and cycle period, respectively. DHZ means the distance from the center of habitable zone. $\sigma(\Delta x$) and $\sigma(\Delta y$) present the standard deviations of astrometric jitter in X(Y) direction. efficiency is the detection efficiency of each star. The symbol * represents large range of activity level.} The unmarked parameters, including $T_{eff}$, log$g$, radius and mass, are from \citet{2021Paegert}. The rotation and cycle periods are from Sect. \ref{s32}. References for other parameter are the following:(1) \citet{1984Noyes}; (2) \citet{1988Perrin}; (3) \citet{1991Duncan}; (4) \citet{1995Baliunas}; (5) \citet{1996Henry}; (6) \citet{1999Allende}; (7) \citet{2004Wright}; (8) \citet{2006Luck}; (9) \citet{2006Gray}; (10) \citet{2008Mamajek}; (11) \citet{2009Holmberg}; (12) \citet{2010Cvetkovic}; (13) \citet{2010Isaacson}; (14) \citet{2011Casagrande}; (15) \citet{2013Pace}; (16) \citet{2013Carretta}; (17) \citet{2013Ramrez}; (18) \citet{2015Paletou}; (19) \citet{2015Jofr}; (20) \citet{2015Terrien}; (21) \citet{2017Theia}; (22) \citet{2017Luck}; (23) \citet{2018Boro}; (24) \citet{2018Aguilera}; (25) \citet{2019Houdebine}; (26) \citet{2019Hojjatpanah}; (27) \citet{2019Stassun}; (28) \citet{2020Steinmetz}; (29) \citet{2020Reiners}; (30) \citet{2021Gomes}; (31) \citet{2021Hirsch}; (32) \citet{2021Perdigon}; (33) \citet{2021Latkovi}; (34) \citet{2022Soubiran}; (35) \citet{2022Chulkov}; (36) \citet{2023Yang}; (37) \citet{2023Hardegree}; }

\end{deluxetable*}
\end{longrotatetable}

\bibliography{ms}{}
\bibliographystyle{aasjournal}

\listofchanges

\end{document}